\documentclass[preprintnumbers,nofootinbib,a4paper,superscriptaddress]{revtex4-1}

\usepackage{amsmath,latexsym,amssymb,amsfonts}
\usepackage{amsthm}
\usepackage{float}
\usepackage{graphicx}
\usepackage{epstopdf}
\usepackage{caption}
\usepackage{subfigure}
\usepackage{bm}
\usepackage{xcolor}
\usepackage{hyperref}
\usepackage[dvips]{psfrag}
\usepackage{tikz-cd}
\usepackage{soul}

\setstcolor{red}

\theoremstyle{definition}

\begin{document}

\title{Holomorphic Conformal Cartan Geometry:\\ Conformal Gravity, Twistors, and Ambitwistors}

\author{Chen-Hsu Chien}
\email[E-mail: ]{chien@math.muni.cz}
\affiliation{Department of Mathematics and Statistics, Masaryk University}

\begin{abstract}
Four-dimensional conformal gravity, twistor theory, and ambitwistor theory are reviewed within the unified framework of holomorphic conformal Cartan geometry. Starting from the spinorial representation of the geometry, the normal $\mathfrak{sl}(4,\mathbb{C})$-valued Cartan connection, its curvature $2$-form, and the associated tractor bundles are systematically formulated. Evaluated on the real Lorentzian slice, the quadratic curvature action is shown to reproduce four-dimensional Weyl--Bach conformal gravity upon fixing a Weyl gauge. Local twistor and dual twistor bundles are constructed as associated spin tractor bundles, illustrating how parallel transport yields the classical incidence relations in flat space and chiral integrability constraints in curved backgrounds: anti-self-duality for $\alpha$-surfaces and self-duality for $\beta$-surfaces. Furthermore, by pairing local twistors and dual twistors via an invariant canonical pairing, ambitwistor space is shown to parameterize complex null geodesics and symmetrically accommodate both left- and right-handed Weyl curvatures. This framework establishes a direct gauge-theoretic bridge linking conformal Cartan geometry to twistor and ambitwistor theories.
\end{abstract}

\maketitle

\section{Introduction}

Conformal gravity has attracted sustained interest as a scale-invariant, gauge-theoretic extension of Einstein's general relativity. Defined by the four-dimensional quadratic action built from the Weyl curvature $2$-form, $S_{\text{CG}} \propto \int_M \mathbf{W} \wedge \star \mathbf{W}$ \cite{Weyl1918, Bach1921}, the theory possesses distinct field-theoretic features: it is power-counting renormalizable in four dimensions, admits a gauge formulation under the conformal group $\mathrm{SO}(4,2)$ and its spin cover $\mathrm{SU}(2,2)$ \cite{KakuTownsend1977, Mannheim1989, Wheeler1991}, and arises naturally within holographic boundary configurations \cite{Maldacena2011}. Over the past decades, differential geometric formulations have clarified the geometric foundations of conformal gravity, establishing that the Bach tensor field equations arise directly from the Yang--Mills-type equations of the normal conformal Cartan connection \cite{Merkulov1984, KorzynskiLewandowski2003, AttardFrancois2015}.

In parallel with gauge-theoretic formulations of gravitation, Penrose's twistor program \cite{Penrose1967, PenroseRindler1986} was developed to reformulate spacetime geometry in terms of complex projective geometry, exploiting the intrinsic conformal invariance of massless fields and null rays. In flat spacetime, the incidence relation establishes a canonical double fibration between complexified Minkowski space $\mathbb{C}\mathbb{M}$ and projective twistor space $\mathbb{P}\mathbb{T}$. In curved backgrounds, Penrose's \textit{non-linear graviton construction} \cite{Penrose1976, Ward1977, WardWells1990, MasonWoodhouse1996} proved that anti-self-dual (ASD) vacuum spacetimes---characterized by vanishing right-handed Weyl curvature ($\mathbf{W}_R = 0$)---correspond bijectively to holomorphic deformations of twistor space. The underlying geometric mechanism relies on the Frobenius integrability of two-dimensional complex $\alpha$-surfaces.

However, extending the non-linear graviton construction to generic non-chiral spacetimes with simultaneously non-vanishing left-handed ($\mathbf{W}_L$) and right-handed ($\mathbf{W}_R$) Weyl curvatures encounters a fundamental structural barrier, historically designated as the \textbf{googly problem} \cite{Penrose1999}. Because generic curved spacetimes admit neither integrable $\alpha$-planes nor $\beta$-planes, reconstructing non-anti-self-dual geometries directly from deformations of a single projective twistor space $\mathbb{P}\mathbb{T}$ remains notoriously difficult. Penrose argued that this chiral asymmetry should not be remedied by symmetric ad hoc constructions, but rather through a quantum-geometric paradigm where the dual twistor acts as a non-commutative differential operator on twistor space, $W_\alpha \leftrightarrow \hbar \partial / \partial Z^\alpha$, culminating in his \textit{palatial twistor theory} \cite{Penrose2015}. 

A parallel geometric resolution is provided by \textbf{ambitwistor theory} \cite{Isenberg1978, Witten1978, LeBrun1983, LeBrun1985}. By pairing primary and dual twistors subject to the constraint $W_\alpha Z^\alpha = 0$, projective ambitwistor space $\mathbb{P}\mathbb{A}$ parameterizes complex null geodesics, symmetrically accommodating both left- and right-handed Weyl curvatures ($\mathbf{W}_L, \mathbf{W}_R$) within a single contact manifold. However, employing classical ambitwistors to reconstruct gravitational metrics encounters severe obstructions: unthickened complex deformations of $\mathbb{P}\mathbb{A}$ capture only the conformal geometry, naturally yielding four-derivative conformal gravity ($B_{ab} = 0$) rather than Einstein gravity. Restricting to vacuum Einstein metrics ($R_{ab} = 0$) requires extending the complex structure to the third formal neighborhood $\mathbb{A}^{(3)}$ inside $\mathbb{P}\mathbb{T} \times \mathbb{P}\mathbb{T}^*$ \cite{LeBrun1983, LeBrun1991}. In practice, this extension faces non-trivial cohomological obstructions and reduces to standard spacetime perturbation theory, forfeiting the non-perturbative solvability of the original twistor program.

This limitation motivated a paradigm shift from classical geometric reconstruction to worldsheet quantum field theory in \textbf{ambitwistor string theory} \cite{MasonSkinner2014, Berkovits2004, AdamoMason2014}. By quantizing $\mathbb{P}\mathbb{A}$ as the target space of a chiral two-dimensional conformal field theory, worldsheet path integrals localize directly onto the solutions of the Cachazo--He--Yuan (CHY) scattering equations \cite{CachazoHeYuan2014}. This yields an efficient formulation of tree-level gauge and gravitational scattering amplitudes, while naturally clarifying why the critical low-energy spectrum of unthickened ambitwistor models governs conformal gravity.

A natural, rigorous mathematical framework for the geometric foundation of conformal gravity, twistor, and ambitwistor theories is provided by \textbf{Cartan geometry} and parabolic tractor calculus \cite{Cartan1923, Kobayashi1972, BastonEastwood1989, Sharpe1997, Cap2005, CapSlovak2009, CurryGover2014}. In this context, local twistors emerge directly from tractor calculus in the spin representation of holomorphic conformal Cartan geometry \cite{Dighton1974, BaileyEastwoodGover1994, AttardFrancois2016_1, AttardFrancois2016_2}.

A holomorphic conformal Cartan geometry $(\mathcal{G}, \mathcal{A})$ of type $(G,H)$ over a complex $4$-manifold $M$ consists of a principal $H$-bundle $\mathcal{G} \to M$ endowed with a Cartan connection $\mathcal{A} \in \Omega^1(\mathcal{G}, \mathfrak{g})$, where $G = \mathrm{SL}(4,\mathbb{C})$, $\mathfrak{g} = \mathfrak{sl}(4,\mathbb{C})$, and $H \subset G$ is the parabolic subgroup stabilizing a totally isotropic two-dimensional subspace in $\mathbb{C}^4$, satisfying:
\begin{itemize}
    \item $R_h^* \mathcal{A} = \mathrm{Ad}_{h^{-1}} \mathcal{A} \quad \forall h \in H$,
    \item $\mathcal{A}(X^\dagger) = X \in \mathfrak{h}$ for every fundamental vertical vector field $X^\dagger \in \Gamma(V\mathcal{G})$ generated by $X \in \mathfrak{h}$,
    \item $\mathcal{A}_p : T_p \mathcal{G} \xrightarrow{\cong} \mathfrak{g}$ is a linear vector space isomorphism at every point $p \in \mathcal{G}$.
\end{itemize}
Because $G$ is semisimple and $H$ is a parabolic subgroup associated with the $|1|$-grading $\mathfrak{g} = \mathfrak{g}_{-1} \oplus \mathfrak{g}_0 \oplus \mathfrak{g}_1$, the pair $(\mathcal{G}, \mathcal{A})$ endows $M$ with a parabolic geometry \cite{Cap2005, CapSlovak2009}. Specifically, the quotient bundle $\mathfrak{g}/\mathfrak{h} \cong \mathfrak{g}_{-1}$ induces a complexified conformal structure $[g]$ on $M$, where the center $\mathbb{C}^* \subset G_0$ of the Levi subgroup $G_0 \subset H$ parameterizes local conformal rescalings \cite{CurryGover2014}. Under this structure, the local twistor bundle naturally emerges as the associated tractor bundle $\mathbb{T} = \mathcal{G} \times_H \mathbb{C}^4$ \cite{Dighton1974, BaileyEastwoodGover1994, Cap2005, CapSlovak2009, AttardFrancois2016_1, AttardFrancois2016_2}, with $\mathcal{A}$ governing parallel transport across $M$.

The structure of this review is organized as follows. In Sec.~\ref{sec:hccg}, the spinorial representation of normal holomorphic conformal Cartan geometry is reviewed, and the associated local twistor and dual twistor bundles are constructed. In Sec.~\ref{sec:cg}, the derivation of the four-dimensional Weyl--Bach conformal gravity action from the quadratic trace of the Cartan curvature is reviewed under the choice of a Weyl gauge. In Sec.~\ref{sec:tt}, twistor theory is formulated within the local twistor bundle framework, reviewing the flat-space incidence relations and the chiral integrability conditions on curved $\alpha$-surfaces. Finally, in Sec.~\ref{sec:at}, ambitwistor theory is reviewed by symmetrically pairing the local twistor and dual twistor bundles, analyzing how the space of null geodesics accommodates the complete non-chiral curvature structure.

\section{Spin Representation of Holomorphic Conformal Cartan Geometry}\label{sec:hccg}

Complexified conformal spacetime is canonically modeled on the compactified Klein space $\mathbb{C}\mathbb{M}^c \cong G/H$, where $G = \mathrm{SL}(4,\mathbb{C})$ and $H \subset G$ is the parabolic subgroup stabilizing a two-dimensional subspace in $\mathbb{C}^4$. A holomorphic conformal Cartan geometry of type $(G, H)$ over a complex $4$-manifold $M$ equips spacetime with a complex conformal structure $[g]$---defined as an equivalence class of non-degenerate holomorphic metrics modulo local complex Weyl rescalings $g \sim \Omega^2(x) g$ with $\Omega(x) \in \mathbb{C}^*$. On flat complexified Minkowski space $\mathbb{C}\mathbb{M}$, a canonical representative metric is given by the holomorphic line element $ds^2 = \det(dx^{AA'}) = \frac{1}{2}\epsilon_{AB}\epsilon_{A'B'} dx^{AA'} dx^{BB'}$.\footnote{Under the standard Pauli matrix isomorphism, complexified coordinates are represented by $x^{AA'} := x^a \sigma_a^{AA'} = \begin{pmatrix} x^0 + x^3 & x^1 - ix^2 \\ x^1 + ix^2 & x^0 - x^3 \end{pmatrix}$, where $x \in \mathbb{C}^4$ (reducing to $\mathbb{R}^4$ on the real Minkowski slice).} This holomorphic formulation algebraically decouples the unprimed ($S$) and primed ($S'$) chiral spin bundles, enabling the Levi factor $G_0 \subset H$ to act via independent left- and right-handed transformations $\mathrm{SL}(2,\mathbb{C})_L \times \mathrm{SL}(2,\mathbb{C})_R \times \mathbb{C}^*$.

The underlying principal bundle $\mathcal{G} \to M$ has structure group $H$. In the spinorial block representation, $H$ is realized as the lower block-triangular subgroup decomposing as the semidirect product $H = G_0 \ltimes G_1$:\footnote{The semidirect product structure guarantees that the unipotent radical $G_1$ is normal in $H$. Conjugation by the Levi factor $G_0$ preserves $G_1$ via $g_0 g_1 g_0^{-1} \in G_1$ for all $g_0 \in G_0$ and $g_1 \in G_1$, ensuring that special conformal transformations transform covariantly under Lorentz rotations and complex dilations.}
\begin{equation}
H = \left\{ \begin{pmatrix} a^{1/2} C_L & 0 \\ -i b & a^{-1/2} C_R \end{pmatrix} \;\middle|\; a \in \mathbb{C}^*, \, C_{L,R} \in \mathrm{SL}(2,\mathbb{C}), \, b \in \mathbb{C}^{2 \times 2} \right\} \subset \mathrm{SL}(4,\mathbb{C}).
\end{equation}

Infinitesimally, the Lie algebra $\mathfrak{g} = \mathfrak{sl}(4,\mathbb{C})$ admits a strict $|1|$-grading $\mathfrak{g} = \mathfrak{g}_{-1} \oplus \mathfrak{g}_0 \oplus \mathfrak{g}_1$, parameterized in block form as:\footnote{Mathematical literature on parabolic geometry typically places the negative-degree subspace $\mathfrak{g}_{-1}$ in the lower-left block; the present convention follows the standard physics and twistor literature.}
\begin{equation}
\mathfrak{g} = \left\{ \underbrace{\begin{pmatrix} 0 & i \rho \\ 0 & 0 \end{pmatrix}}_{\mathfrak{g}_{-1}} + \underbrace{\begin{pmatrix} \mathcal{C}_L + \frac{1}{2}\alpha \mathbb{I}_2 & 0 \\ 0 & \mathcal{C}_R - \frac{1}{2}\alpha \mathbb{I}_2 \end{pmatrix}}_{\mathfrak{g}_{0}} + \underbrace{\begin{pmatrix} 0 & 0 \\ -i \beta & 0 \end{pmatrix}}_{\mathfrak{g}_{1}} \;\middle|\; \alpha \in \mathbb{C}, \, \mathcal{C}_{L,R} \in \mathfrak{sl}(2,\mathbb{C}), \, \rho, \beta \in \mathbb{C}^{2 \times 2} \right\}.
\end{equation}
This $|1|$-grading defines the parabolic subalgebra $\mathfrak{h} = \mathfrak{g}_0 \oplus \mathfrak{g}_1$. The grade-zero subspace constitutes the reductive Levi subalgebra $\mathfrak{g}_0 \cong \mathfrak{sl}(2,\mathbb{C})_L \oplus \mathfrak{sl}(2,\mathbb{C})_R \oplus \mathbb{C}$, spanned by the chiral complexified Lorentz algebras and the one-dimensional center generated by the grading element (representing complex dilations). The abelian nilpotent radical $\mathfrak{g}_1$ and its negative-degree counterpart $\mathfrak{g}_{-1}$ represent complexified special conformal transformations and translations in bispinor form, respectively.

\vspace{0.3cm}
\paragraph*{\textbf{Holomorphic Conformal Cartan Connection and Curvature}}
The geometry is governed by a holomorphic conformal Cartan connection $\mathcal{A} \in \Omega^1(\mathcal{G}, \mathfrak{sl}(4,\mathbb{C}))$ and its curvature $2$-form $\mathcal{F} = d\mathcal{A} + \mathcal{A} \wedge \mathcal{A} \in \Omega^2(\mathcal{G}, \mathfrak{sl}(4,\mathbb{C}))$. In the spinorial block representation, these forms are parameterized as:
\begin{equation}
\mathcal{A} = \begin{pmatrix} \Gamma_L + \frac{1}{2} a \mathbb{I}_2 & i\Theta \\ -iP & \Gamma_R - \frac{1}{2} a \mathbb{I}_2 \end{pmatrix}, \quad
\mathcal{F} = \begin{pmatrix} W_L + \frac{1}{2} f \mathbb{I}_2 & iT \\ -iC & W_R - \frac{1}{2} f \mathbb{I}_2 \end{pmatrix},
\end{equation}
where $\Gamma_{L,R}$ denote the chiral spin connections, $a$ is the dilation $1$-form, $\Theta$ is the soldering form, and $P$ is the $1$-form governing special conformal transformations. The curvature components decompose into the chiral Weyl $2$-forms $W_L = R_L + (\Theta \wedge P)_0$ and $W_R = R_R + (P \wedge \Theta)_0$, where $(\cdot)_0$ denotes the trace-free projection and $R_{L,R} = d\Gamma_{L,R} + \Gamma_{L,R} \wedge \Gamma_{L,R}$ are the spin curvatures. The remaining components comprise the dilation curvature $f = da + \mathrm{Tr}(\Theta \wedge P)$, the conformal torsion $T = D\Theta + a \wedge \Theta$, and the grade-one curvature $C = DP - a \wedge P$, where $D$ denotes the exterior covariant derivative with respect to $\Gamma_{L,R}$. Strictly speaking, local gauge fields on spacetime are obtained by pulling back the principal bundle connection along a local section $\sigma: U \subset M \to \mathcal{G}$ via $A = \sigma^* \mathcal{A}$; following standard practice in theoretical physics, the explicit pullback map $\sigma^*$ is suppressed to avoid notational clutter, denoting the local spacetime connection and curvature directly by $\mathcal{A}$ and $\mathcal{F}$.

\vspace{0.3cm}
\paragraph*{\textbf{Gauge Transformations}}
Under a local gauge transformation $g: U \to H$ (corresponding to a change of section $\sigma \mapsto \sigma g$), the connection and curvature transform via:
\begin{equation}
\mathcal{A} \mapsto \mathcal{A}^g = g^{-1} \mathcal{A} g + g^{-1} dg, \quad \mathcal{F} \mapsto \mathcal{F}^g = g^{-1} \mathcal{F} g.
\end{equation}
Restricting $g$ to the Levi subgroup $G_0 \subset H$ with parameterization $g = \mathrm{diag}(\Omega^{1/2}\, g_L, \Omega^{-1/2}\, g_R)$ for $\Omega \in \mathbb{C}^*$ and $g_{L,R} \in \mathrm{SL}(2,\mathbb{C})$, the constituent forms transform according to their conformal weights:
\begin{equation}
\Theta \mapsto \Omega^{-1}\, g_L^{-1} \,\Theta\, g_R, \quad T \mapsto \Omega^{-1}\, g_L^{-1}\, T \,g_R, \quad P \mapsto \Omega \,g_R^{-1}\, P \,g_L, \quad C \mapsto \Omega \,g_R^{-1}\, C \,g_L.
\end{equation}
The dilation $1$-form transforms affinely as $a \mapsto a + d(\ln \Omega)$, leaving the dilation curvature $f$ invariant. The trace-free curvatures transform tensorially under the Lorentz subgroup ($W_{L,R} \mapsto g_{L,R}^{-1}\, W_{L,R}\, g_{L,R}$). Consequently, the vanishing of torsion ($T=0$) and the vanishing of Weyl curvature ($W_{L,R}=0$) are gauge-invariant geometric conditions.

\vspace{0.3cm}
\paragraph*{\textbf{Normality Condition}}
To eliminate algebraic gauge redundancies (unphysical degrees of freedom), the standard Kostant normality condition $\partial^* \mathcal{F} = 0$ is imposed \cite{CapSlovak2009, AttardFrancois2016_1}. This constraint forces the conformal torsion to vanish ($\mathbf{T} = 0$), uniquely determining the spin connections $\mathbf{\Gamma}_{L,R}(\mathbf{\Theta}, a)$ in terms of the soldering form $\mathbf{\Theta}$ and dilation connection $a$. It further identifies the grade-one connection $P$ with the Schouten $1$-form $\mathbf{P}$. The algebraic symmetry of the Schouten tensor ensures $\mathrm{Tr}(\mathbf{\Theta} \wedge \mathbf{P}) = 0$, reducing the dilation curvature to $\mathbf{f} = da$. Under this reduction, the grade-one curvature $C$ becomes the Cotton $2$-form $\mathbf{C}$, and the trace-free grade-zero curvatures $W_{L,R}$ become the chiral Weyl $2$-forms $\mathbf{W}_{L,R}$.

The normal holomorphic conformal Cartan connection $\mathcal{A}$ and its curvature $\mathcal{F}$ take the canonical form:
\begin{equation} \label{CC}
\mathcal{A} = \begin{pmatrix} \mathbf{\Gamma}_L + \frac{1}{2} a \mathbb{I}_2 & i\mathbf{\Theta} \\ -i\mathbf{P} & \mathbf{\Gamma}_R - \frac{1}{2} a \mathbb{I}_2 \end{pmatrix}, \quad
\mathcal{F} = \begin{pmatrix} \mathbf{W}_L + \frac{1}{2} \mathbf{f} \mathbb{I}_2 & 0 \\ -i\mathbf{C} & \mathbf{W}_R - \frac{1}{2} \mathbf{f} \mathbb{I}_2 \end{pmatrix},
\end{equation}
where $\mathbf{W}_L = d\mathbf{\Gamma}_L + \mathbf{\Gamma}_L \wedge \mathbf{\Gamma}_L + \mathbf{\Theta} \wedge \mathbf{P}$, $\mathbf{W}_R = d\mathbf{\Gamma}_R + \mathbf{\Gamma}_R \wedge \mathbf{\Gamma}_R + \mathbf{P} \wedge \mathbf{\Theta}$, and $\mathbf{C} = d\mathbf{P} + \mathbf{P} \wedge \mathbf{\Gamma}_L + \mathbf{\Gamma}_R \wedge \mathbf{P} - a \wedge \mathbf{P}$.

\vspace{0.3cm}
\paragraph*{\textbf{Local Twistor and Dual Local Twistor Bundles}}
The Cartan connection canonically defines a covariant derivative on the associated vector bundle $\mathbb{T} = \mathcal{G} \times_H \mathbb{C}^4$, identified as the \textbf{local twistor bundle} (or spin tractor bundle) \cite{Cap2005, CapSlovak2009, Dighton1974, BaileyEastwoodGover1994, AttardFrancois2016_2}. Geometrically, $\mathbb{T}$ possesses a canonical filtration $0 \to (S')^* \to \mathbb{T} \to S \to 0$. Because the unipotent radical $G_1 \subset H$ acts non-trivially on the unprimed spinor subspace, fixing a Weyl structure (or metric gauge) splits the bundle into $\mathbb{T} \cong S \oplus (S')^*$, allowing smooth sections $Z \in \Gamma(\mathbb{T})$ to be represented as column vectors $Z^\alpha = (\omega^A, \pi_{A'})^T$. Parallel transport in $\mathbb{T}$ is governed by the local twistor connection $\nabla^{\mathbb{T}} = d + \mathcal{A}$:
\begin{equation} \label{TC}
\nabla^{\mathbb{T}} \begin{pmatrix} \omega \\ \pi \end{pmatrix} = d \begin{pmatrix} \omega \\ \pi \end{pmatrix} + \begin{pmatrix} \mathbf{\Gamma}_L + \frac{1}{2} a \mathbb{I}_2 & i\mathbf{\Theta} \\ -i\mathbf{P} & \mathbf{\Gamma}_R - \frac{1}{2} a \mathbb{I}_2 \end{pmatrix} \begin{pmatrix} \omega \\ \pi \end{pmatrix} = \begin{pmatrix} \nabla_L^{\mathbb{T}}\,\omega + i\mathbf{\Theta}\pi \\ \nabla_R^{\mathbb{T}}\,\pi - i\mathbf{P}\omega \end{pmatrix}.
\end{equation}
The curvature acts on local twistors via $(\nabla^{\mathbb{T}})^2 Z = \mathcal{F} Z$.

Similarly, the dual vector bundle $\mathbb{T}^* = \mathcal{G} \times_H (\mathbb{C}^4)^*$ defines the \textbf{dual local twistor bundle}. Smooth sections $W \in \Gamma(\mathbb{T}^*)$ are represented by row vectors $W_\alpha = (\lambda_A, \mu^{A'}) \in \Gamma(S^* \oplus S')$. Parallel transport is governed by the dual connection $\nabla^{\mathbb{T}^*} W = dW - W\mathcal{A}$:
\begin{equation} \label{DTC}
\nabla^{\mathbb{T}^*} \begin{pmatrix} \lambda & \mu \end{pmatrix} = d \begin{pmatrix} \lambda & \mu \end{pmatrix} - \begin{pmatrix} \lambda & \mu \end{pmatrix} \begin{pmatrix} \mathbf{\Gamma}_L + \frac{1}{2} a \mathbb{I}_2 & i\mathbf{\Theta} \\ -i\mathbf{P} & \mathbf{\Gamma}_R - \frac{1}{2} a \mathbb{I}_2 \end{pmatrix} = \begin{pmatrix} \nabla_L^{\mathbb{T}^*} \lambda + i\mu\mathbf{P} & \nabla_R^{\mathbb{T}^*} \mu - i\lambda\mathbf{\Theta} \end{pmatrix},
\end{equation}
with curvature action $(\nabla^{\mathbb{T}^*})^2 W = -W \mathcal{F}$.

\vspace{0.3cm}
\paragraph*{\textbf{Global and Holonomy Twistors}}
A \textbf{global twistor} (resp. \textbf{dual twistor}) is defined as a globally parallel section satisfying $\nabla^{\mathbb{T}} Z = 0$ (resp. $\nabla^{\mathbb{T}^*} W = 0$). Path-independent global solutions exist on $M$ if and only if the underlying manifold is conformally flat ($\mathcal{F} = 0$).

When curvature is non-zero ($\mathcal{F} \neq 0$), integrability is obstructed. Sections can instead be parallel transported along complex null curves $\gamma(\tau)$ with tangent vector $K$. \textbf{Holonomy twistors} and \textbf{holonomy dual twistors} are sections $Z \in \Gamma(\gamma^* \mathbb{T})$ and $W \in \Gamma(\gamma^* \mathbb{T}^*)$ of the pullback bundles over $\gamma$ satisfying the directional transport equations $\nabla_K^{\mathbb{T}} Z = 0$ and $\nabla_K^{\mathbb{T}^*} W = 0$.\footnote{In parts of the spinorial and twistor literature, the directional covariant derivative along a curve is also denoted by $\not\!\nabla$, $\nabla_K$, or $\frac{D}{d\tau}$.}

\vspace{0.3cm}
\paragraph*{\textbf{Real Minkowski Spacetime Slice}}
Restricting to Lorentzian signature $(1,3)$ reduces the structure group to the real form $G_M = \mathrm{SU}(2,2)$ and the parabolic subgroup to $H_M \subset \mathrm{SU}(2,2)$. The algebraic reality conditions of $\mathfrak{su}(2,2)$ enforce $\mathbf{\Gamma}_R = -\mathbf{\Gamma}_L^\dagger$, implying $\mathbf{W}_R = -\mathbf{W}_L^\dagger$. The dilation $1$-form $a$ and its curvature $\mathbf{f}$ become real-valued, while the soldering form $\mathbf{\Theta}$ and Schouten tensor $\mathbf{P}$ become Hermitian matrix-valued $1$-forms, recovering standard real conformal geometry.

\section{Conformal Gravity from Conformal Cartan Geometry}\label{sec:cg}

The dynamics of conformal gravity can be formulated within the Yang--Mills gauge-theoretic paradigm applied to the real slice of the normal conformal Cartan geometry \cite{KakuTownsend1977, Wheeler1991, Merkulov1984, KorzynskiLewandowski2003, AttardFrancois2015}. Over a four-dimensional pseudo-Riemannian manifold $(M, [g])$ with Lorentzian signature $(1,3)$, the structure group reduces to the real form $G_M = \mathrm{SU}(2,2)$. The gauge-invariant action functional $S[\mathcal{A}_M]$ is defined by the quadratic trace of the normal Cartan curvature $2$-form $\mathcal{F}_M \in \Omega^2(M, \mathfrak{su}(2,2))$:
\begin{equation}
S[\mathcal{A}_M] = -\frac{1}{4g_{\mathrm{CG}}^2} \int_M \mathrm{Tr}\left( \mathcal{F}_M \wedge \star \mathcal{F}_M \right),
\end{equation}
where $g_{\mathrm{CG}}$ is the dimensionless conformal gravity coupling constant and $\star$ denotes the Hodge star operator associated with the conformal class $[g]$.\footnote{In four dimensions, the Hodge star acting on $2$-forms is conformally invariant ($\star_{\Omega^2 g} = \star_g$), rendering the action functional $S[\mathcal{A}_M]$ strictly independent of the representative metric chosen from $[g]$.}

The algebraic structure of the normal Cartan curvature $\mathcal{F}_M$ simplifies the evaluation of the trace. Although the lower-left block contains the Cotton $2$-form $\mathbf{C}$, the vanishing of the conformal torsion ($\mathbf{T} = 0$) in the upper-right block ensures that off-diagonal terms do not contribute to the matrix trace. Moreover, because the chiral Weyl $2$-forms are trace-free ($\mathrm{Tr}(\mathbf{W}_L) = \mathrm{Tr}(\mathbf{W}_R) = 0$), cross-terms with the dilation curvature vanish identically. Taking the trace over the $2 \times 2$ blocks yields:
\begin{equation}
S_{\mathrm{CG}} = -\frac{1}{4g_{\mathrm{CG}}^2} \int_M \left[ \mathrm{Tr}(\mathbf{W}_L \wedge \star \mathbf{W}_L) + \mathrm{Tr}(\mathbf{W}_R \wedge \star \mathbf{W}_R) + \mathbf{f} \wedge \star \mathbf{f} \right].
\end{equation}

To recover the standard metric formulation of conformal gravity, local scale invariance is broken by imposing the \textbf{Weyl gauge} ($a = 0$, setting $\mathbf{f} = 0$), which reduces $\mathbf{\Gamma}_{L,R}(\mathbf{\Theta}, 0)$ to the Levi-Civita spin connection of the chosen metric $g \in [g]$. In this gauge, the chiral sum reproduces the classical Weyl action $\frac{1}{2} \int_M W_{abcd} W^{abcd} \mathrm{vol}_g$, and varying with respect to $\mathbf{\Theta}$ yields the Bach tensor field equations $B_{ab} = 0$ \cite{Bach1921, KorzynskiLewandowski2003}. While formulated here in the $\mathrm{SU}(2,2)$ spin cover for direct compatibility with twistors, this dynamical reduction is completely equivalent to the fundamental $\mathrm{SO}(4,2)$ conformal formulation \cite{Wheeler1991, AdamoMason2014, KorzynskiLewandowski2003}.

\section{Twistor Theory from Holomorphic Conformal Cartan Geometry}\label{sec:tt}

The differential geometric bridge between holomorphic conformal Cartan geometry and twistor theory is established by identifying twistors as parallel sections of the local twistor bundle $\mathbb{T}$ \cite{PenroseRindler1986, Cap2005, Dighton1974, BaileyEastwoodGover1994, AttardFrancois2016_2}. In a conformally flat background ($\mathcal{F} = 0$), parallel transport is path-independent, giving rise to global twistors $Z \in \Gamma(\mathbb{T}) \cong \mathbb{C}^4$. In a curved background ($\mathcal{F} \neq 0$), non-zero curvature obstructs global integrability, restricting solutions to path-dependent holonomy twistors $Z \in \Gamma(\gamma^* \mathbb{T})$ along null curves $\gamma$. As shown by Penrose \cite{Penrose1976, Ward1977, WardWells1990, MasonWoodhouse1996}, reconstructing a global curved twistor space requires the integrability of complex two-dimensional $\alpha$-surfaces, which is sensitive exclusively to left-handed (anti-self-dual) gravitational degrees of freedom.

\subsection{Global Twistors in a Flat Background}

In a conformally flat background, the Cartan curvature vanishes identically ($\mathcal{F} = 0$), rendering the local twistor connection flat and integrable. Consequently, a parallel section is uniquely determined everywhere by its value at any single reference point. The vector space of universal, path-independent global twistors $Z \in \Gamma(\mathbb{T})$ forms a four-dimensional complex vector space canonically isomorphic to $\mathbb{C}^4$.

In a flat vacuum background with Cartesian gauge choice ($\mathbf{\Gamma}_{L,R} = 0$, $a = 0$, $\mathbf{P} = 0$, and $\mathbf{\Theta}^{AA'} = dx^{AA'}$), the connection $\mathcal{A}$ is simplified significantly. Evaluating the parallel transport equation $\nabla^{\mathbb{T}} Z = 0$ along translation directions $dx^{AA'}$ yields:
\begin{equation}
    \begin{cases} 
        \partial_{AA'}\omega^B + i\delta^B_A \pi_{A'} = 0 \\ 
        \partial_{AA'}\pi_{B'} = 0 
    \end{cases}
    \quad\implies\quad 
    \begin{cases} 
        \omega^A(x) = \omega_0^A - i x^{AA'}\pi_{0A'} = -i(x^{AA'} + i y_0^{AA'})\pi_{0A'} \\ 
        \pi_{A'}(x) = \pi_{0A'} 
    \end{cases}
\end{equation}
where $\omega_0^A$ and $\pi_{0A'}$ are constant integration spinors, and $\omega_0^A = y_0^{AA'}\pi_{0A'}$ is parameterized by a complex displacement vector $y_0^{AA'}$. This recovers Penrose's classical \textbf{incidence relation} directly from parallel transport of twistors.

The geometric correspondence encoded by the incidence relation is formalized via the \textbf{Penrose double fibration} \cite{Penrose1967, PenroseRindler1986, BastonEastwood1989, Cap2005}. The correspondence space (flag manifold) $\mathbb{F} = \mathbb{F}_{1,2}(\mathbb{C}^4)$ is defined as the space of incident pairs $(x, [Z]) \in \mathbb{C}\mathbb{M} \times \mathbb{P}\mathbb{T}$. Because the incidence relation algebraically fixes $\omega^A$ given $x^{AA'}$ and $\pi_{A'}$, the manifold $\mathbb{F}$ is canonically trivialized as $\mathbb{F} \cong \mathbb{C}\mathbb{M} \times \mathbb{C}\mathbb{P}^1$, equipped with two canonical projections:
\begin{equation}
    \begin{tikzcd}
        & \mathbb{F} \arrow[dl, "\rho_1"'] \arrow[dr, "\zeta_1"] & \\
        \mathbb{P}\mathbb{T} & & \mathbb{C}\mathbb{M}
    \end{tikzcd}
\end{equation}
Coordinatized by $(x^{AA'}, [\pi_{A'}])$, the spacetime projection $\zeta_1: \mathbb{F} \to \mathbb{C}\mathbb{M}$ given by $\zeta_1(x^{AA'}, [\pi_{A'}]) = x^{AA'}$ fibers $\mathbb{F}$ with Riemann spheres $\zeta_1^{-1}(x) \cong \mathbb{C}\mathbb{P}^1$. Conversely, the twistor projection $\rho_1: \mathbb{F} \to \mathbb{P}\mathbb{T}$ maps to projective twistor space via $\rho_1(x^{AA'}, [\pi_{A'}]) = [\omega^A(x), \pi_{A'}]$, where $\omega^A = -i(x^{AA'} + i y_0^{AA'})\pi_{0A'}$. The fibers $\rho_1^{-1}([Z]) \cong \mathbb{C}^2$ define totally null $2$-dimensional \textbf{$\alpha$-planes}.\footnote{Subtracting the incidence relations for two points $x_1, x_2$ on an $\alpha$-plane yields $\Delta x^{AA'} \pi_{A'} = 0$, implying $\Delta x^{AA'} = \xi^A \pi^{A'}$. Thus $\det(\Delta x^{AA'}) = 0$, showing that an $\alpha$-plane is a totally null $2$-surface.} This double fibration establishes the fundamental twistor duality: \textit{a point in complexified spacetime corresponds to a projective line in twistor space, while a point in projective twistor space corresponds to a null $\alpha$-plane in complexified spacetime $\mathbb{C}\mathbb{M}$.}

Restricting to the real Minkowski slice $\mathbb{M}$ (where $\mathbf{x}^{AA'} = \bar{\mathbf{x}}^{AA'}$), twistor space $\mathbb{T}$ inherits the $\mathrm{SU}(2,2)$-invariant Hermitian form:
\begin{equation}
    \langle Z, Z \rangle = Z^\dagger \Omega Z = \begin{pmatrix} \bar{\omega}^{A'} & \bar{\pi}_A \end{pmatrix} \begin{pmatrix} 0 & \mathbb{I}_2 \\ \mathbb{I}_2 & 0 \end{pmatrix} \begin{pmatrix} \omega^A \\ \pi_{A'} \end{pmatrix} = \bar{\omega}^{A'}\pi_{A'} + \bar{\pi}_A \omega^A = 2s,
\end{equation}
where $s$ denotes the quantum helicity. 

For zero quantum helicity ($s = 0$), corresponding to $\omega_0^A = 0$, the incidence relation reduces to $\omega^A(\mathbf{x}) = -i \mathbf{x}^{AA'} \pi_{0A'}$. Restricting the Penrose double fibration to the real Minkowski slice yields the real correspondence structure $\mathbb{P}\mathbb{N} \leftarrow \mathbb{F}_{\mathbb{M}} \rightarrow \mathbb{M}$, where $\mathbb{F}_{\mathbb{M}} \cong \mathbb{M} \times \mathbb{C}\mathbb{P}^1$ and $\mathbb{P}\mathbb{N} \subset \mathbb{P}\mathbb{T}$ denotes the null twistor hypersurface ($\langle Z, Z \rangle = 0$). Under this mapping, a real spacetime point $\mathbf{x} \in \mathbb{M}$ corresponds to a projective line lying entirely within $\mathbb{P}\mathbb{N}$. \textit{However, a generic point in $\mathbb{P}\mathbb{T}$ with non-zero norm lies outside $\mathbb{P}\mathbb{N}$, defining an $\alpha$-plane in $\mathbb{C}\mathbb{M}$ that does not intersect the real Minkowski slice $\mathbb{M}$.}

For non-zero quantum helicity ($s \neq 0$), with $\omega_0^A \neq 0$, the incidence relation $\omega^A(\mathbf{x}) = -i (\mathbf{x}^{AA'} + i y_0^{AA'})\pi_{0A'}$ yields the norm $\langle Z, Z \rangle = 2 y_0^{AA'} \bar{\pi}_A \pi_{0A'} = 2s$, revealing helicity as a topological invariant associated with a complex spatial displacement $y_0^{AA'}$. The null hypersurface $\mathbb{P}\mathbb{N}$ divides projective twistor space into top and bottom regions $\mathbb{P}\mathbb{T}^\pm$. In this regime, a projective line in $\mathbb{P}\mathbb{T}^\pm$ corresponds in real spacetime $\mathbb{M}$ to a \textbf{Robinson congruence}---a shear-free, twisting family of null geodesics.

\subsection{Holonomy Twistors in a Curved Background}

In a curved spacetime ($\mathcal{F} \neq 0$), non-zero curvature precludes the existence of globally parallel twistors. Twistor sections are therefore restricted to path-dependent holonomy twistors $Z \in \Gamma(\gamma^* \mathbb{T})$ along curves $\gamma$. To reconstruct a geometric twistor space in the presence of curvature, one evaluates the integrability of two-dimensional complex surfaces swept out by null vectors, known as curved $\alpha$-surfaces.

Let $\gamma(\tau)$ be a complex null curve with tangent vector $K$. The covariant derivative along $K$ is given by:
\begin{equation}
    \nabla^{\mathbb{T}}_K \begin{pmatrix} \omega^A \\ \pi_{A'} \end{pmatrix} = \begin{pmatrix} (\nabla_L)_K + \frac{1}{2} (K \lrcorner a) \mathbb{I}_2 & i (K \lrcorner \mathbf{\Theta}) \\ -i (K \lrcorner \mathbf{P}) & (\nabla_R)_K - \frac{1}{2} (K \lrcorner a) \mathbb{I}_2 \end{pmatrix} \begin{pmatrix} \omega^A \\ \pi_{A'} \end{pmatrix} = 0.
\end{equation}
In a gauge where $a = 0$ and $\mathbf{P} = 0$, parameterizing an arbitrary null tangent vector along an $\alpha$-surface as $K^{AA'} = \xi^A \pi^{A'}$ for an arbitrary unprimed spinor $\xi^A$, the off-diagonal contraction vanishes: $(K \lrcorner \mathbf{\Theta}^{AA'})\pi_{A'} = \xi^A \pi^{A'} \pi_{A'} = 0$ due to spinor skew-symmetry. The directional transport equations decouple into:
\begin{equation}
    (\nabla_L)_K\, \omega^A = 0 \quad \text{and} \quad (\nabla_R)_K\, \pi_{A'} = 0.
\end{equation}

For these directional solutions to sweep out an integrable $\alpha$-surface $\mathbf{\Sigma}$, the integrability condition requires the pullback of the Cartan curvature to vanish ($\mathcal{F}|_{\mathbf{\Sigma}} = 0$). In the spinor basis, this pullback takes the form:
\begin{equation}
    \mathcal{F}|_{\mathbf{\Sigma}} = \begin{pmatrix} \mathbf{W}_L{}^A{}_{BCD}\Sigma^{CD} & 0 \\ -i(\mathbf{C}_L{}_{A'BCD} \Sigma^{CD} + \mathbf{C}_R{}_{A'BC'D'} \tilde{\Sigma}^{C'D'}) & \mathbf{W}_R{}_{A'}{}^{B'}{}_{C'D'} \tilde{\Sigma}^{C'D'} \end{pmatrix},
\end{equation}
where $\Sigma^{AB} = \epsilon_{C'D'} ( \mathbf{\Theta}^{AC'} \wedge \mathbf{\Theta}^{BD'} )$ spans the anti-self-dual (ASD) $2$-forms and $\tilde{\Sigma}^{A'B'} = \epsilon_{CD} ( \mathbf{\Theta}^{CA'} \wedge \mathbf{\Theta}^{DB'} )$ spans the self-dual (SD) $2$-forms. Tangent vectors $V = \xi_1^A \pi^{A'} \partial_{AA'}$ and $U = \xi_2^A \pi^{A'} \partial_{AA'}$ spanning an $\alpha$-surface satisfy $\Sigma^{AB}(V, U) = 0$ identically due to $\epsilon_{C'D'} \pi^{C'} \pi^{D'} = 0$, whereas $\tilde{\Sigma}^{A'B'}(V, U) \neq 0$. Consequently, while the left-handed Weyl curvature $\mathbf{W}_L$ drops out of the restriction, the vanishing of $\mathcal{F}|_{\mathbf{\Sigma}}$ strictly requires the right-handed curvature components to vanish ($\mathbf{W}_R = 0$ and $\mathbf{C}_R = 0$), forcing $\nabla_R$ to be pure gauge. The existence of integrable $\alpha$-surfaces therefore requires the spacetime to be \textbf{anti-self-dual (ASD)}, precisely aligning with Penrose's non-linear graviton construction.

When $\mathbf{W}_R = 0$, Penrose's non-linear graviton construction \cite{Penrose1976, Ward1977, WardWells1990, MasonWoodhouse1996} builds the curved projective twistor space $\mathcal{PT}$ by gluing local patches $U$ and $\tilde{U}$. Using local coordinates $Z^\alpha = (\omega^A, \pi_{A'})^T$ and $\tilde{Z}^\alpha = (\tilde{\omega}^A, \tilde{\pi}_{A'})^T$, the constancy of $\pi_{A'}$ across $\alpha$-surfaces trivializes the primed sector ($\tilde{\pi}_{A'} = \pi_{A'}$), restricting the holomorphic transition functions (which encode the left-handed Weyl curvature $\mathbf{W}_L$) entirely to the unprimed sector:
\begin{equation}
    \tilde{\omega}^A = \omega^A + \varepsilon F^A(\omega^B, \pi_{B'}), \quad \text{with} \quad F_A = \frac{\partial H}{\partial \omega^A},
\end{equation}
where $H \in H^1(\mathcal{PT}, \mathcal{O}(2))$ is a homogeneous Hamiltonian generating a symplectomorphism that preserves the fiberwise holomorphic volume form $d\omega^A \wedge d\omega_A = d\tilde{\omega}^A \wedge d\tilde{\omega}_A$.

\section{Ambitwistor Theory from Holomorphic Conformal Cartan Geometry}\label{sec:at}

While Penrose's non-linear graviton construction describes anti-self-dual (left-handed) vacuum spacetimes through deformations of projective twistor space $\mathbb{P}\mathbb{T}$, generic physical spacetimes require the simultaneous presence of both left-handed ($\mathbf{W}_L$) and right-handed ($\mathbf{W}_R$) gravitational degrees of freedom. Accommodating both chiralities within a unified holomorphic twistor framework is historically connected to the googly problem \cite{Penrose1999, Penrose2015}. Rather than relying exclusively on chiral $\alpha$-planes, a non-chiral geometric description is provided by \textbf{ambitwistor theory} \cite{Isenberg1978, Witten1978, LeBrun1983, LeBrun1985, MasonSkinner2014}, which parameterizes the space of complex null geodesics. Within holomorphic conformal Cartan geometry, an ambitwistor is defined as a pair $(Z^\alpha, W_\alpha) \in \Gamma(\mathbb{T} \times \mathbb{T}^*)$ consisting of a primary local twistor and a dual local twistor. The fundamental \textbf{ambitwistor space} $\mathbb{A}$ is defined by:
\begin{equation}
    \mathbb{A} = \left\{ (Z^\alpha, W_\alpha) \in \mathbb{T} \times \mathbb{T}^* \;\middle|\; W_\alpha Z^\alpha = \text{constant} \right\}.
\end{equation}
By the Leibniz rule, the covariant derivative of the ambitwistor pairing vanishes identically: $\nabla^{\mathbb{A}} (W_\alpha Z^\alpha) := (\nabla^{\mathbb{T}^*} W_\alpha) Z^\alpha + W_\alpha (\nabla^{\mathbb{T}} Z^\alpha) = 0$, ensuring that the pairing is a parallel scalar invariant under the Cartan connection.

\subsection{Global Ambitwistors in a Flat Background}

A global primary twistor $Z^\alpha = (\omega^A, \pi_{A'})^T \in \Gamma(\mathbb{T})$ and a global dual twistor $W_\alpha = (\lambda_A, \mu^{A'}) \in \Gamma(\mathbb{T}^*)$ satisfy the global parallel transport equations $\nabla^{\mathbb{T}}Z^\alpha = 0$ and $\nabla^{\mathbb{T}^*}W_\alpha = 0$, respectively. In a conformally flat ($\mathcal{F} = 0$) and simply connected background, the local twistor and dual local twistor connections are flat, yielding four-dimensional vector spaces of global parallel sections $\Gamma(\mathbb{T}) \cong \mathbb{C}^4$ and $\Gamma(\mathbb{T}^*) \cong (\mathbb{C}^4)^*$, respectively. In this flat regime, evaluating parallel transport in a standard Cartesian gauge generates the paired primary and dual \textbf{incidence relations}:
\begin{equation}
    \begin{cases} 
        \omega^A(x) = \omega_0^A - i x^{AA'}\pi_{0A'} = -i (x^{AA'} + i y_\alpha^{AA'}) \pi_{0A'} \qquad\text{with}\qquad \pi_{A'}(x) = \pi_{0A'} \\ 
        \mu^{A'}(x) = \mu_0^{A'} + i x^{AA'}\lambda_{0A} = i (x^{AA'} - i y_\beta^{AA'}) \lambda_{0A} \qquad\text{with}\qquad \lambda_A(x) = \lambda_{0A}
    \end{cases}
\end{equation}
where $\omega_0^A = y_\alpha^{AA'}\pi_{0A'}$ and $\mu_0^{A'} = y_\beta^{AA'}\lambda_{0A}$ are parameterized by constant complex translation vectors $y_\alpha^{AA'}$ and $y_\beta^{AA'}$. The primary incidence relation defines an $\alpha$-plane in $\mathbb{C}\mathbb{M}$, while the dual incidence relation defines a conjugate $\beta$-plane.

The geometric correspondence between complexified spacetime $\mathbb{C}\mathbb{M}$ and projective ambitwistor space $\mathbb{P}\mathbb{A}$ is formalized via the \textbf{ambitwistor double fibration} \cite{LeBrun1983, MasonSkinner2014, Cap2005}:
\begin{equation}
    \begin{tikzcd}
        & \mathbb{F}_{\mathbb{A}} \arrow[dl, "\rho_2"'] \arrow[dr, "\zeta_2"] & \\
        \mathbb{P}\mathbb{A} & & \mathbb{C}\mathbb{M}
    \end{tikzcd}
\end{equation}
The correspondence space $\mathbb{F}_{\mathbb{A}}$ consists of mutually incident point-twistor triplets $(x^{AA'}, [Z^\alpha], [W_\alpha])$. Parameterized by $(x^{AA'}, [\pi_{A'}], [\lambda_A])$, the spacetime projection $\zeta_2: \mathbb{F}_{\mathbb{A}} \to \mathbb{C}\mathbb{M}$ given by $\zeta_2(x^{AA'}, [\pi_{A'}], [\lambda_A]) = x^{AA'}$ fibers $\mathbb{F}_{\mathbb{A}}$ with products of Riemann spheres $\zeta_2^{-1}(x) \cong \mathbb{C}\mathbb{P}^1 \times \mathbb{C}\mathbb{P}^1$. Conversely, the ambitwistor projection $\rho_2: \mathbb{F}_{\mathbb{A}} \to \mathbb{P}\mathbb{A}$ maps $(x^{AA'}, [\pi_{A'}], [\lambda_A]) \mapsto ([Z^\alpha(x, \pi)], [W_\alpha(x, \lambda)])$. The fiber $\rho_2^{-1}([Z^\alpha], [W_\alpha])$ represents the intersection of the corresponding $\alpha$-plane and $\beta$-plane in $\mathbb{C}\mathbb{M}$, which forms a one-dimensional complex null geodesic $\gamma \subset \mathbb{C}\mathbb{M}$. Under this correspondence, a point in spacetime corresponds to a quadric $\mathbb{C}\mathbb{P}^1 \times \mathbb{C}\mathbb{P}^1 \subset \mathbb{P}\mathbb{A}$, while a point in projective null ambitwistor space $\mathbb{P}\mathbb{A}_0$ corresponds to a complex null geodesic in $\mathbb{C}\mathbb{M}$.

Evaluating the invariant pairing $W_\alpha Z^\alpha$ across complexified spacetime shows that the position coordinate $x^{AA'}$ cancels identically, yielding a position-independent invariant:
\begin{equation}
    W_\alpha Z^\alpha = \lambda_A(x)\omega^A(x) + \mu^{A'}(x)\pi_{A'}(x) = \lambda_{0A} \omega_0^A + \mu_0^{A'} \pi_{0A'} = 2s.
\end{equation}
For null ambitwistors ($s = 0$, with $\omega_0^A = \mu_0^{A'} = 0$), the primary and dual incidence relations reduce to $\omega^A(x) = -i x^{AA'} \pi_{0A'}$ and $\mu^{A'}(x) = i x^{AA'}\lambda_{0A}$. The tangent distributions of the $\alpha$-plane ($\Delta x^{AA'} = \xi^A \pi_0^{A'}$) and conjugate $\beta$-plane ($\Delta x^{AA'} = \lambda_0^A \eta^{A'}$), spanned by arbitrary spinors $\xi^A$ and $\eta^{A'}$ respectively, intersect along the shared null vector $K^{AA'} = \lambda_0^A \pi_0^{A'}$, defining a unique complex null geodesic $\gamma(\tau) \subset \mathbb{C}\mathbb{M}$. For non-vanishing helicity ($s \neq 0$), the complex displacement vectors yield $W_\alpha Z^\alpha = (y_\alpha^{AA'} + y_\beta^{AA'}) \lambda_{0A} \pi_{0A'} = 2s$. In this case, the null ray is shifted into the complex domain, and its projection onto the real Minkowski slice $\mathbb{M}$ (where $W_\alpha = \bar{Z}_\alpha$) generates a \textbf{Robinson congruence}.

\subsection{Holonomy Ambitwistors in a Curved Background}

In a curved background ($\mathcal{F} \neq 0$), parallel transport is path-dependent. A holonomy ambitwistor consists of a pair $(Z^\alpha, W_\alpha)$ parallel-transported along a complex null geodesic $\gamma(\tau)$ with tangent vector field $K^{AA'} = \lambda^A \pi^{A'}$. The directional parallel transport equations along $\gamma$, $\nabla_K^{\mathbb{T}} Z^\alpha = 0$ and $\nabla_K^{\mathbb{T}^*} W_\alpha = 0$, expand explicitly to:
\begin{equation}
    \begin{cases}
        \left[ (\nabla_L)_K + \frac{1}{2}(K \lrcorner a) \right] \omega^A + i (K \lrcorner \mathbf{\Theta}^{AA'}) \pi_{A'} = 0 \quad\text{and}\quad \left[ (\nabla_R)_K - \frac{1}{2}(K \lrcorner a) \right] \pi_{A'} - i (K \lrcorner \mathbf{P}_{AA'}) \omega^A = 0 \\ 
        \left[ (\nabla_L^*)_K - \frac{1}{2}(K \lrcorner a) \right] \lambda_A + i \mu^{A'} (K \lrcorner \mathbf{P}_{AA'}) = 0 \quad\text{and}\quad \left[ (\nabla_R^*)_K + \frac{1}{2}(K \lrcorner a) \right] \mu^{A'} - i \lambda_A (K \lrcorner \mathbf{\Theta}^{AA'}) = 0.
    \end{cases}
\end{equation}
In a gauge where $a = 0$ and $\mathbf{P} = 0$, the off-diagonal contractions vanish identically along the null direction: $(K \lrcorner \mathbf{\Theta}^{AA'})\pi_{A'} = \lambda^A \pi^{A'} \pi_{A'} = 0$ and $\lambda_A(K \lrcorner \mathbf{\Theta}^{AA'}) = \lambda_A \lambda^A \pi^{A'} = 0$ due to spinor skew-symmetry. The system decouples into four directional equations:
\begin{equation} 
    \begin{cases} 
        (\nabla_L)_K\, \omega^A = 0 \quad\text{and}\quad (\nabla_R)_K\, \pi_{A'} = 0 \\ 
        (\nabla_L^*)_K\, \lambda_A = 0 \quad\text{and}\quad (\nabla_R^*)_K\, \mu^{A'} = 0.
    \end{cases}
\end{equation}
Along the one-dimensional null curve $\gamma$, since $\pi_{A'}$ and $\lambda_A$ are covariantly constant momentum spinors, the directional connections $(\nabla_R)_K$ and $(\nabla_L^*)_K$ are pure gauge along the ray.

The non-chiral structure of ambitwistor space is captured by decomposing the normal Cartan curvature into its anti-self-dual and self-dual components, $\mathcal{F} = \mathcal{F}_{\mathrm{ASD}} + \mathcal{F}_{\mathrm{SD}} \in \Omega^2_{\mathrm{ASD}}(M, \mathfrak{sl}(4,\mathbb{C})) \oplus \Omega^2_{\mathrm{SD}}(M, \mathfrak{sl}(4,\mathbb{C}))$:
\begin{equation}
\mathcal{F}^{\mathbb{T}} \equiv \mathcal{F}_{\mathrm{ASD}} = \begin{pmatrix} \mathbf{W}_L{}^A{}_{BCD} \Sigma^{CD} & 0 \\ -i\mathbf{C}_L{}_{A'BCD} \Sigma^{CD} & 0 \end{pmatrix},
\quad
\mathcal{F}^{\mathbb{T}^*} \equiv \mathcal{F}_{\mathrm{SD}} = \begin{pmatrix} 0 & 0 \\ -i\mathbf{C}_R{}_{A'BC'D'} \tilde{\Sigma}^{C'D'} & \mathbf{W}_R{}_{A'}{}^{B'}{}_{C'D'} \tilde{\Sigma}^{C'D'} \end{pmatrix}.
\end{equation}
Consequently, the left-handed Weyl curvature $\mathbf{W}_L \subset \mathcal{F}^{\mathbb{T}}$ generates holomorphic deformations of the primary position spinor $\omega^A$, while the right-handed Weyl curvature $\mathbf{W}_R \subset \mathcal{F}^{\mathbb{T}^*}$ symmetrically deforms the dual position spinor $\mu^{A'}$.

These paired non-linear deformations generate contact transformations that preserve the canonical holomorphic contact $1$-form on curved projective ambitwistor space $\mathcal{PA}$:
\begin{equation}
    \theta = \frac{1}{2}\left( Z^\alpha dW_\alpha - W_\alpha dZ^\alpha \right) = \frac{1}{2}\left( \tilde{Z}^\alpha d\tilde{W}_\alpha - \tilde{W}_\alpha d\tilde{Z}^\alpha \right).
\end{equation}
Consequently, rather than requiring either $\mathbf{W}_R = 0$ or $\mathbf{W}_L = 0$ to construct an integrable space of surfaces, the ambitwistor bundle formulation symmetrically accommodates both left- and right-handed curvature components $(\mathbf{W}_L, \mathbf{W}_R)$ within a single contact manifold structure \cite{LeBrun1983, LeBrun1985, MasonSkinner2014}.

From the perspective of holomorphic conformal Cartan geometry, the normal Cartan connection $\mathcal{A}$ and its curvature $\mathcal{F}$ provide the geometric kinematics required to formulate both pure gravitational theories and matter couplings. Guided by the double fibration correspondence, dynamical actions can be formulated equivalently on spacetime $M$, curved projective twistor space $\mathcal{PT}$ (for chiral sectors), or curved projective ambitwistor space $\mathcal{PA}$ (for full non-chiral theories) \cite{Mason2005, Berkovits2004, AdamoMason2014, MasonSkinner2014}. On (ambi)twistor space, these functionals are naturally realized as holomorphic BF or Chern--Simons-type actions. Under the Penrose transform and the non-linear Ward--Isenberg--Yasskin--Green correspondence \cite{Ward1977, Isenberg1978, Witten1978}, performing fiberwise integration over the correspondence space $\mathbb{F}_{\mathbb{A}}$ directly reduces the holomorphic ambitwistor action to the four-dimensional quadratic Weyl--Bach conformal gravity action on spacetime $M$ \cite{Mason2005, AdamoMason2014, Adamo2013}. Modern developments have further extended these twistor and ambitwistor action principles to compute conformal and Einstein gravitational scattering amplitudes directly from four-dimensional ambitwistor string models \cite{GeyerLipsteinMason2014}.

\section{Conclusion}\label{sec:conclusion}

In this review, holomorphic conformal Cartan geometry has been presented as a unified bundle-theoretic framework connecting gauge-theoretic conformal gravity, twistor theory, and ambitwistor theory. Infinitesimally modeled on the flat Klein geometry $(\mathrm{SL}(4,\mathbb{C}), H)$ with compactified model space $\mathbb{C}\mathbb{M}^c \cong \mathrm{SL}(4,\mathbb{C})/H$, this geometric formulation provides a direct realization of local scale and conformal symmetries on complexified spacetime.

When evaluated on the real Lorentzian slice with structure group $\mathrm{SU}(2,2)$, the Yang--Mills-type action formed from the quadratic trace of the normal Cartan curvature $2$-form decouples into chiral Weyl and dilation kinetic terms. Imposing the Weyl gauge sets the dilation connection to zero and reduces the normal Cartan connection to the Levi-Civita spin connection, where Euler--Lagrange variation with respect to the soldering form yields the trace-free Bach tensor field equations of four-dimensional Weyl--Bach conformal gravity.

Within this geometric setting, twistors emerge as parallel sections of the associated local tractor bundle $\mathbb{T}$. In flat spacetime, parallel transport generates the classical incidence relation, establishing the Penrose double fibration correspondence with complexified Minkowski space. In curved backgrounds, global integrability is obstructed by non-zero Cartan curvature. Evaluating surface integrability on two-dimensional complex $\alpha$-surfaces requires the vanishing of the right-handed curvature components ($\mathbf{W}_R = 0, \mathbf{C}_R = 0$), recovering Penrose's chiral non-linear graviton construction of curved projective twistor space $\mathcal{PT}$ for anti-self-dual spacetimes.

To accommodate generic non-chiral gravitational fields, ambitwistor space is constructed by pairing the primary and dual local twistor bundles under the invariant canonical pairing constraint $W_\alpha Z^\alpha = 2s$. Because ambitwistors parameterize one-dimensional complex null geodesics rather than two-dimensional chiral surfaces, parallel transport along null directions is governed by directional ordinary differential equations that remain integrable on generic curved backgrounds. In this paired geometry, the primary bundle $\mathbb{T}$ isolates the left-handed Weyl curvature $\mathbf{W}_L$ (deforming the unprimed sector $\omega^A$), while the dual bundle $\mathbb{T}^*$ isolates the right-handed Weyl curvature $\mathbf{W}_R$ (deforming the primed sector $\mu^{A'}$). These deformations are symmetrically combined by preserving the canonical holomorphic contact $1$-form $\theta = \frac{1}{2}(Z^\alpha dW_\alpha - W_\alpha dZ^\alpha)$ on curved projective ambitwistor space $\mathcal{PA}$, thereby incorporating both gravitational chiralities within a single geometric framework.

By systematically bridging holomorphic conformal Cartan geometry with the spin representation of associated tractor bundles, this formulation provides a clear, unified geometric foundation linking conformal gravity to twistor and ambitwistor theories.

\begin{acknowledgments}
The author was supported by the Czech Science Foundation (GA\v{C}R) under the research project ``Cartan supergeometries and Higher Cartan geometries'' (Grant No.~GA24-10887S) and by the European Union's Horizon Europe research and innovation programme under the Marie Sk\l{}odowska-Curie grant agreement No.~101086123 (CaLIGOLA). This research was also supported by the Simons Foundation grant SFI-MPS-T-Institutes-00010825, and by State Treasury funds as part of a task commissioned by the Minister of Science and Higher Education under the project ``Organization of the Simons Semesters at the Banach Center --- New Energies in 2026--2028'' (MNiSW/2025/DAP/491).
\end{acknowledgments}

\end{document}